\documentclass[sigconf,screen,nonacm]{acmart}
\AtBeginDocument{%
  }

\setcopyright{cc}
\setcctype[4.0]{by}

\begin{document}

\title{Design Concept: Scaffolding Geopolitical Reflection Among Tech Workers}

\author{Sydney Reis}
\email{Sydney.Reis@cs.ox.ac.uk}
\affiliation{%
  \institution{Responsible Technology Institute, Department of Computer Science, University of Oxford}
  \city{Oxford}
  \country{UK}
}

\renewcommand{\shortauthors}{Reis}

\begin{abstract}

This paper presents a speculative Human-Computer Interaction design proposal for encouraging geopolitical reflexivity amongst tech workers at geopolitically relevant technology companies. Recent scholarship in International Relations and Science and Technology Studies increasingly recognizes technology firms and their workers as geopolitical actors whose decisions shape international dynamics. However, existing Responsible Innovation and Responsible AI approaches rarely engage with the geopolitical narratives and imaginaries that underpin contemporary AI development.

Building upon RI scholarship on reflexivity, reflective HCI, and creative HCI work on computational narratives, this paper proposes an AI-enabled interactive narrative system in which users engage with a speculative scenario centred on technology, power, and geopolitics. Through narrative interaction, archetype assignment, and socially scaffolded workshop reflection, the system aims to encourage target users to critically examine their assumptions, values, and positionality within broader sociotechnical systems. We argue that speculative narrative systems may offer a productive avenue for introducing geopolitical reflexivity into responsible technology initiatives without relying on prescriptive or moralising approaches. 

\end{abstract}


\maketitle

\textbf{Reference:}\newline
Sydney Reis. 2026. Design Concept: Scaffolding Geopolitical Reflection Among Tech Workers. In \textit{Proceedings of The First Reflection in Creative Experience (RiCE) Workshop (RiCE W1)}. ACM Creativity \& Cognition 2026, London, UK.

\section{Introduction}

Recent scholarship in International Relations examines how powerful firms and their workers in the AI stack are developing geopolitical agency\cite{schwartz_exclusive_2025, lehdonvirta_compute_2024, gill_whoever_2020, pavel_ai_2023, chesterman_silicon_2025, srivastava_ai_2024, bomont_challenging_2025, quansah_deepseek_2025, srivastava_algorithmic_2023, nadibaidze_startups_2025, csernatoni_myth_2025}. While this scholarship often has its roots in the literature of international political economy\cite{strange_retreat_1996, farrell_weaponized_2019}, it is also inspired by recent and current events at the intersection of technology, power, and geopolitics. For instance, Elon Musk's remote satellite company Starlink has, at different points during Russia's most recent invasion of Ukraine, unilaterally granted and denied connectivity access to Ukrainian and Russian forces during active conflict\cite{roulette_musk_2025}. Venture capital-funded technology startups raise capital based on promises to establish sovereign jurisdictions in the likes of Greenland and Honduras\cite{palencia_honduran_2022, bernstein_who_2023}. OpenAI's Stargate project promises to launch "OpenAI for Countries" whereby states can build their private and public AI infrastructure atop the OpenAI stack\cite{openai_introducing_nodate}. 

One common thread running through each of these examples is the shared narrative of technological development and use without accountability or oversight. Terms such as \textit{democratic AI}, \textit{strategic competition}, and \textit{technological race} are invoked by technology companies and leaders to justify technological advancement\cite{fischer_technological_2021}, often at the expense of ethical, safe, or secure AI development and/or robust AI governance.\cite{walter_managing_2024, schmid_arms_2025, the_white_house_executie_office_winning_2025, vance_remarks_2025, depirri_tale_2025, wei_how_2024, the_white_house_initial_2025, csernatoni_eus_2025}. The first International AI Safety Report acknowledges the "strong competitive pressure[s] which may lead [governments and AI companies] to de-prioritize risk management"\cite{bengio_international_nodate}. In this context, firms accountable primarily to private interests, rather than voters in democratic countries, are increasingly shaping dynamics within the international system. 

Tech workers within these companies have previously demonstrated awareness of their role in geopolitics, often through how the technologies developed and provided by their workplaces are used in global conflict\cite{wakabayashi_google_2018, anonymous_google_and_amazon_workers_we_2021}. More recent examples of tech worker resistance to geopolitical events within tech companies have also been documented, particularly in media reporting\cite{tan_unlikely_2024, reis_ai_2025, di_navigating_2025, kelly_palantir_2026}. Science and Technology Studies (STS) has long held that technologies take on the values of the people who create them\cite{jasanoff_dreamscapes_2015}. If this premise is accepted, then tech workers, especially at those companies that are developing geopolitically relevant technologies, become a necessary intervention site to encourage the responsible actions that reduce harmful global impacts\cite{sharma_ais_2024}. 

Narratives are a significant part of how firms engage with the current geopolitical system\cite{csernatoni_myth_2025, jasanoff_dreamscapes_2015}. From an STS perspective, firm narratives and the ideologies that underpin them\footnote{this concept is explored more completely by Sheila Jasanoff and Sang Hyun-Kim as "sociotechnical imaginaries" in their work \textit{Dreamscapes of Modernity}} contribute to the design logics of the systems and technologies, and are acted upon from both within and without firms. Firms therefore create, and in turn, are created by, popular narratives such as \textit{strategic competition}, \textit{technological race}, \textit{democratic AI}, and various uses of the term \textit{sovereignty}. Interrogating the narratives may be a crucial first step in encouraging tech workers to think through the how, why, and global impacts of the technologies that they work on; which can contribute to more ethical AI from the ground up.  

This design proposal seeks to explore the tensions between responsible AI development, geopolitics, tech workers, and the narratives that fuel AI development today, building upon existing work in Human-Computer Interaction (HCI) and creative HCI. I present a tool targeted towards tech workers that interrogates geopolitically relevant narratives and the history and sociology of technology to encourage reflexivity related to their positionality and potential agency in the current geopolitical moment. This goal is aligned with the broader goals of RAI, which advocates for better development and design choices from the ground up\cite{schuurbiers_what_2011, stoyanovich_using_2024, goni_tooling_2024, gold_virginia_2021}. At the moment, geopolitical reflexivity is missing from RAI. Existing work on reflection tools in creative HCI can offer productive ways to engage tech workers who are often ideologically motivated and have little time for reflection in the workplace. 

\section{Background and Motivation}

\subsection{Inspiration}

AI companies across the stack are gaining geopolitical influence and agency\cite{farrell_weaponized_2019, lehdonvirta_compute_2024, gill_whoever_2020, pavel_ai_2023, chesterman_silicon_2025, srivastava_ai_2024, bomont_challenging_2025, quansah_deepseek_2025, srivastava_algorithmic_2023, nadibaidze_startups_2025}. With the introduction of AI as a general purpose technology\cite{crafts_artificial_2021}, the subsequent consolidation of power that AI has brought frontier labs\cite{hao_empire_2025}, and the geopolitically relevant narratives that drive AI development today\cite{csernatoni_myth_2025}, the people that build, sell, market, and make decisions about AI ought to be aware of the global impacts of their work\cite{reis_ai_2025}. Considering the geopolitical relevance of AI companies, tech workers within these companies should become key sites of intervention for RAI efforts. While the relative agency of these workers may be variable, they are positioned to make decisions - whether of technical, strategic, or business in nature - that have ramifications for how states interact with one another in the international system\cite{khan_ai_2022, farrell_weaponized_2019, tan_unlikely_2024, wei_how_2024}. However, the intersection of geopolitics and technology is rarely addressed in Responsible Innovation (RI) or Responsible AI (RAI) initiatives\footnote{While RI and RAI are distinct but related traditions, both broadly seek to encourage more socially responsive technological development}\cite{possati_exploring_2024}.

In their foundational paper Stilgoe et al. introduce four "pillars" of RI\cite{stilgoe_developing_2013}. These are anticipation, reflexivity, inclusion, and responsiveness. This design proposal focuses especially on the \textit{reflexivity}\footnote{Defined by Stilgoe et al. as "building actors’ and institutions’ reflexivity means rethinking prevailing conceptions about the moral division of labour within science and innovation\cite{swierstra_exploring_2009}. Reflexivity directly challenges assumptions of scientific amorality and agnosticism. Reflexivity asks scientists, in public, to blur the boundary between their role responsibilities and wider, moral responsibilities. It therefore demands openness and leadership within cultures of science and innovation."} pillar of RI. Grimpe et al. later expanded and applied Stilgoe et al.'s four pillars to HCI\cite{grimpe_towards_2014}. Here, Grimpe et al. connect reflexivity to Sch\"{o}n's notion of "reflection-in-action"\cite{schon_reflective_2016} and argue that "Technology is not conceptualized as a product to be designed, delivered and then to be ‘ticked off’, but as an “open space” that different users and other stakeholders continually engage with, and that they may challenge at times"\cite{grimpe_towards_2014}. Therefore, reflexivity offers a promising entry point for engaging tech workers whose views may be shaped by dominant narratives about AI and geopolitics\cite{csernatoni_myth_2025}. While reflexivity alone cannot guarantee more responsible technological outcomes\cite{stilgoe_developing_2013}, if appropriately scaffolded\cite{slovak_tools_2025}, it may serve as a necessary precursor to broader responsible engagement.

There is already a body of work in the RI and HCI literature that engages tech workers on reflection and responsibility through toolkits and systems. This body of work includes the use of cards\cite{elsayed-ali_responsible_2023}, reflective systems\cite{colombo_tech_2026}, and games\cite{harrell_chimeriagrayscale_2018}. Further work in HCI and creative HCI on computational narratives\cite{kreminski_why_2020, kreminski_evaluating_2019-1} and the use of AI for reflection tools is also particularly salient\cite{baumer_reviewing_2014, bentvelzen_revisiting_2022}. Combined, these approaches provide a roadmap for developing a bespoke intervention that encourages tech workers to engage more meaningfully with the geopolitical impacts of their technologies.

\section{Design Concept}

\subsection{Overview}

The proposed tool is an AI system whereby the user engages in an interactive narrative storyline. The tool is intended for tech workers at geopolitically relevant companies, to be used at the workplace, ideally during workshops or other appropriately socially scaffolded sessions\cite{slovak_reflective_2017, slovak_tools_2025}.

\subsection{Assumptions}

The design rationale for this proposal makes several assumptions about the target population of technology workers in geopolitically relevant companies. These include that the target population is willing and able to engage in reflection, though they may have limited time and resources for reflection, and that they may have established beliefs about technology and geopolitics based on dominant narratives at this current moment. Workers, however, may have different levels of agency and autonomy within their organisations to enact change and make more responsible decisions in the workplace. They may also be mission oriented and motivated not just by financial remuneration, but also by a desire to enact positive change through their technologies. However, despite internal motivations, workers may face organisational pressures that make reflection or responsible design socially costly to engage in or discuss, or may be outright discouraged in various forms.

Research in psychology suggests that some people are better than others in engaging in reflection\cite{chaddock_understanding_2014}. Moreover, reflection in HCI tools must be appropriately scaffolded, preferably within social situations, it cannot be assumed to just happen in a vacuum\cite{slovak_reflective_2017}. This design proposal also assumes that reflection will lead to better, safer, fairer, or more responsible technology development, proliferation, and design outcomes\cite{stilgoe_developing_2013, grimpe_towards_2014, fisher_midstream_2006, schuurbiers_what_2011}. However, this type of reflection has not yet been robustly applied to the current geopolitical environment.\cite{possati_exploring_2024}

According to Green and Jenkins, interactive narratives are "stories that allow readers to determine the direction of the plot, often at key decision points".\cite{green_interactive_2014} In their review of interactive narratives, Green and Jenkins found that interactive narratives have been found to "encourage a felt responsibility for a characters actions" therefore making the engagement "more enjoyable" as well as "attitudinal and behavioural change"\cite{green_interactive_2014}. There is also some research to suggest that interactive narratives could lead to value change. Steinemann et al., for instance, found that prosocial behaviour followed from study participants who were engaged and appreciative of the interactive narratives\cite{steinemann_interactive_2017}. Moreover, speculative design in HCI has long been held as an avenue for imagining more just practices, values, and power relations\cite{dunne_speculative_2013, sondergaard_fabulation_2023, blythe_research_2014}.

\subsection{Tensions}

The tool must be engaging for the busy tech worker, such as those working in start-ups, who may have little time and resources to engage in more fundamental level reflection\cite{nadibaidze_startups_2025, ryan_model_2023}. It should not aim to assign correctness to views\cite{koralus_philosophic_2025}. The user must experience high amounts of agency in their engagement with the tool, so that they feel that the reflection is their own and not imposed on them.

\subsection{Experience}

The system would function as a chat style interface that is hosted on an external website, resembling the landing page of existing Large Language Models (LLMs), with space for an added storyline above the chatbot. Using a classic chat style interface would ensure that the platform is familiar and recognisable to the user population. The colour palette would be soft and neutral to evoke a sense of calm in the user. 

The user begins at an introductory screen with an introductory scenario. After receiving background information on the scenario and being instructed to reply to the prompts as closely as possible to their own beliefs, values, and personality, the user is immediately confronted with a choice that requires reflexive engagement with their core values. The scenario is geopolitically relevant but does not invoke real countries or technologies. The system reacts to what the user has inputted with an update to the scenario, and the user responds in turn. Every decision that the user makes results in a different outcome. After sequencing through several of these iterations, the user is assigned an "archetype" that comes with a description and a saveable avatar\cite{jung_archetypes_1980}. The system relies on the use of AI in the back-end, and bespoke digital art on the frontend. Approximately ten archetypes would be developed, which will be inspired by historical technologists and leaders. 

\subsection{Moments and Mechanisms of Reflection}

The key moments of reflection will occur during interaction with the system itself, and afterwards during a guided reflection exercise, preferably in the form of a workshop. The archetype assignment functions as a conversational prompt to encourage further socially-mediated reflection amongst coworkers. Each archetype description will contain unique strengths and weaknesses to guide discussion.

The reflection that occurs during use of the tool is both in-the-moment ("reflection-in-action")\cite{schon_reflective_2016} and retrospective. Reflection may begin during interaction with prompts as the user responds to each of the prompts and sees how the storyline evolves based on their responses to the prompts, but deeper reflection would occur during group discussions and/or workshops centred on the use of the tool\cite{baumer_reviewing_2014, baumer_reflective_2015}. Workshop discussion should be kept to 1 hour maximum, and should include a short presentation on which archetype each participant was assigned, an oral reflection on how they think they got there, whether they agree or disagree with the assignment of the archetype, and whether they wish that they would have been assigned a different archetype. 

This tool would support critical and ethical reflection\cite{sengers_reflective_2005}. By exploring themes such as the role of narratives, the interaction between technology and geopolitics, and the inherent complexity of human interaction with their tools across time and space through story, role-play, ambiguity, and collective discussion, the system aims to encourage users to critically examine their own assumptions, values, and positionality within broader sociotechnical systems. By keeping this tool user-led, anchored in the workplace, and by avoiding the use of buzz words or reference to current events while staying anchored in real-world data, rumination\cite{eikey_beyond_2021} is ideally minimised. 

\subsection{Discussion}

\subsubsection{Opportunities}

This design concept proposes a novel idea to a genuine problem with real world, downstream implications for safe, responsible, and ethical technology development. It engages a user population that is not often engaged, and in many ways, has yet to be appropriately categorized in the literature. The tool attempts to leverage HCI and creative HCI research on AI-mediated reflection, while addressing some of the critiques of responsible innovation toolkits, such as ineffectiveness and failure to engage with critical, root issues\cite{goni_tooling_2024, wong_seeing_2023}.   

\subsubsection{Risks}

There are several anticipated risks to the use of this tool. These include unintended rumination\cite{eikey_beyond_2021} after the assignment of an archetype, the potential for social judgement within the workplace during the workshop element, bias in the data used (the historical and geopolitical data that would underpin the RAG-tuning tends to be written by the victor, after all), and the potential for users to perceive the tool as pushing a particular narrative or political ideology. 

In order to try to mitigate these risks, care should be taken to ensure that the archetypes that are assigned are not inherently hostile or could be interpreted to mean that their assignment can be attributed to personal failing. Each archetype will be assigned strengths and weaknesses, and there will be no assigned hierarchy to archetypes. Efforts will also be made to ensure explainability, where the system explains how and why it adapted the story, and what "decisions" were made in archetype assignment.

\subsubsection{Open Questions}

As the design prototype is developed, there are several open questions for further discussion, such as: 

\begin{itemize}
    \item \textbf{Conceptual}
    \begin{enumerate}
        \item How can reflective systems encourage geopolitical reflexivity without becoming prescriptive or moralising?
    \end{enumerate}

    \item \textbf{Design}
    \begin{enumerate}
        \item What forms of narrative ambiguity best support reflection while preserving user agency and interest?
        \item Can playfulness or aesthetic softness reduce defensiveness in ethically charged reflective systems?
    \end{enumerate}

    \item \textbf{Methodological}
    \begin{enumerate}
        \item How might researchers evaluate whether meaningful geopolitical reflection has occurred?
    \end{enumerate}
\end{itemize}
 
\section{Future Work and Conclusion}

The design of the tool will be informed by ethnographic inquiry, semi-structured interviews, and design workshops with the target population. In line with participatory design approaches, this will allow for a deeper understanding of the needs and beliefs of the target population\cite{gregory_scandinavian_nodate}. 

This design proposal has outlined the design logic of an RI-inspired tool for tech workers at geopolitically relevant companies. Given that tech workers at these companies have increasing levels of geopolitical influence, RI and RAI efforts should attempt to reach this newly relevant population. HCI and creative HCI may offer a unique way to do this, while at the same time addressing some of the critiques of the responsible innovation movement. This design concept proposes the development of an AI-enabled, RAG fine-tuned tool for the target population that encourages reflection on narratives that currently underpin technology and geopolitics and may discourage more responsible, safe, or ethical AI and technology development. By engaging narratives rather than politics, policies, or current events, it is anticipated that the tool will allow for unthreatening reflection amongst a target population that may already have relatively crystallized beliefs.

\bibliographystyle{ACM-Reference-Format}
\bibliography{Library}

\end{document}